\documentclass{aa}  
\usepackage{graphicx}
\usepackage{txfonts}
\usepackage{natbib}
\begin{document}
   \title{A forming disk at z$\sim$0.6:\\ Collapse of a gaseous disk or major merger remnant?}


   \author{M. Puech\inst{1,2}
         \and
         F. Hammer\inst{2}
	  \and
	  H. Flores\inst{2}
	  \and
	  B. Neichel\inst{2}
	  \and
	  Y. Yang\inst{2}
          }

   \offprints{mpuech@eso.org}

   \institute{ESO, Karl-Schwarzschild-Strasse 2, D-85748 Garching bei M\"unchen, Germany
\and 
GEPI, Observatoire de Paris, CNRS, University Paris Diderot; 5 Place Jules Janssen, 92190 Meudon, France
}

   \date{Received...accepted...}

\abstract{Local spiral galaxies contain roughly two-thirds of the
  present-day stellar mass density. However, the formation process of
  disks is still poorly understood.}{We present and analyze
  observations of J033241.88-274853.9 at z=0.6679 to understand how
  its stellar disk was formed.}{We combine multi-wavelength EIS,
  HST/ACS, Spitzer/IRAC, and GALEX imaging with FLAMES/GIRAFFE 3D
  spectroscopy to study its color-morphology and spatially-resolved
  kinematics. A spectral energy distribution (SED) is constructed and
  physical properties extracted using stellar population
  models.}{J033241.88-274853.9 is a blue, young (320$^{+590} _{-260}$
  Myr, 90\% confidence interval) stellar disk embedded in a very
  gas-rich (f$_{gas}$=73-82\% with $\log({M_{stellar}/M_\odot})=9.45
  \pm ^{0.28} _{0.14}$) and turbulent phase that is found to be
  rotating on large spatial scales. We identified two unusual
  properties of J033241.88-274853.9. (1) The spatial distributions of
  the ionized gaseous and young stars show a strong decoupling; while
  almost no stars can be detected in the southern part down to the
  very deep detection limit of ACS/UDF images (accounting for the
  light spread by seeing effects), significant emission from the [OII]
  ionized gas is detected. (2) We detect an excess of velocity
  dispersion in the southern part of J033241.88-274853.9 in comparison
  to expectations from a rotating disk model.}{We considered two disk
  formation scenarios, depending on the gaseous phase geometry. In the
  first one, we examined whether J033241.88-274853.9 could be a young
  rotating disk that has been recently collapsed from a pre-existing,
  very gas-rich rotating disk. This scenario requires two (unknown)
  additional assumptions to explain the decoupling between the
  distribution of stars and gas and the excess of velocity dispersion
  in the same region. In a second scenario, we examine whether
  J033241.88-274853.9 could be a merger remnant of two gas-rich disks.
  In this case, the asymmetry observed between the gas and star
  distributions, as well as the excess of velocity dispersion, find a
  common explanation. Shocks produced during the merger in this region
  can be ionized easily and heat the gas while preventing star
  formation.  This makes this scenario more satisfactory than the
  collapse of a pre-existing, gas-rich rotating disk.}


   \maketitle
%

\section{Introduction}
While the majority of the local stellar mass density is locked into
spiral galaxies, our understanding of the formation of galactic disks
still remains incomplete \citep{mayer08}. It has been suggested that a
large fraction of local disks could have been rebuilt subsequent to
a major merger since z=1, owing to the remarkable coincidence of the
evolution of the merger rate, morphology, and fraction of actively
star-forming galaxies \citep{hammer05}. At first sight, this ``spiral
rebuilding'' scenario might seem inconsistent with numerical
simulations, which often predict that the product of a major merger
between two spirals is an elliptical galaxy. However, it has also been
shown that gas expelled in tidal tails during the merger can be re-accreted and
reform a disk after such an event \citep{barnes02}. More recently,
theoretical and numerical clues to the disk-rebuilding hypothesis
have been accumulated (\citealt{robertson06,lotz08,hopkins08}).

In the framework of an ESO large program called IMAGES
(``Intermediate-MAss Galaxy Evolution Sequence'',
\citealt{ravikumar07,yang08}, hereafter Paper I), we have been
gathering multi-wavelength data on a representative sample of emission
line, intermediate-mass galaxies at z$\sim$0.6, i.e., 6 Gyr ago. These
galaxies are the progenitors of present-day spirals, which contain
approximately two-thirds of the present-day stellar mass density
\citep{hammer07}. In Paper I, we presented the GIRAFFE spatially
resolved kinematics of this sample of 65 galaxies. In \cite{neichel08}
(hereafter Paper II), we studied in detail their color-morphology
using HST/ACS multi-wavelength imaging. In \cite{puech08} (Paper III),
we analyzed their dynamical properties through the evolution of the
near-infrared Tully-Fisher relation. This series of papers has
revealed a surprisingly high fraction of kinematically nonrelaxed
galaxies (Paper I), while well-relaxed spiral rotating disks appear to
be evolving since z$\sim$0.6, i.e., over the past 6 Gyr, by a factor
as much as two, both in number (Paper II) and stellar mass (Paper
III). Clearly, the spiral-disk galaxy history over the last 6 Gyr
seems to be particularly agitated, much more than initially thought.

Space imaging combined with ground-based 3D spectroscopy can reveal
the physical processes at work in such distant galaxies with
unprecedented detail (\citealt{puech07b}). In the present letter, we
present another spin-off of the IMAGES Large Program, which is a very
young ($\sim$ 300 Myr old) rotating disk captured during its formation
process. This article is organized as follows: Sect. 2 presents the
morphological and kinematic properties of J033241.88-274853.9, Sect. 3
its SED fitting, and Sect. 4 the derivation of its stellar mass and
star formation rate. Section 5 presents two possible scenarios of disk
formation. A discussion and conclusion are given in Sect. 6.
Throughout, we adopt $H_0$=70 km/s/Mpc, $\Omega _M$=0.3, and $\Omega
_\Lambda$=0.7, and the $AB$ magnitude system.

\section{Morpho-kinematics}
\begin{figure*}
\centering
\includegraphics[width=18cm]{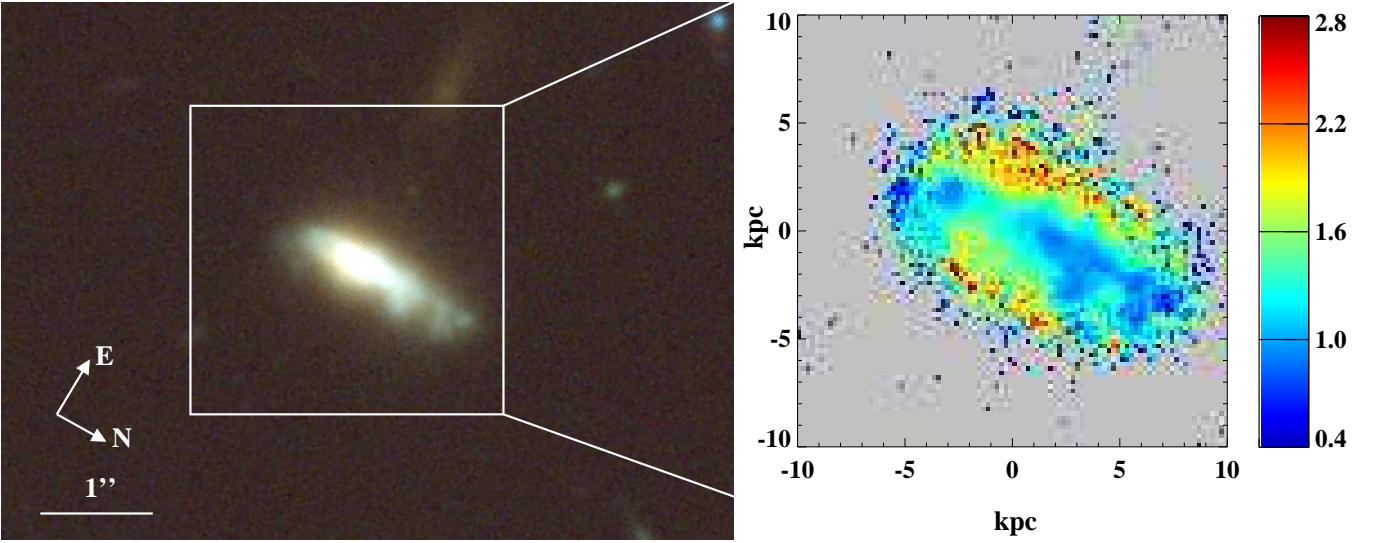}
\caption{\emph{Left:} HST/ACS B-V-z-band image of J033241.88-274853.9
  form the Hubble Ultra-Deep Field (0.03 arcsec/pix,
  \citealt{beckwith06}). \emph{Right:} B-z color map.}
\label{view}
\end{figure*}

Figure \ref{view} shows a large view of J033241.88-274853.9, which
reveals a relatively edge-on Peculiar/Tadpole galaxy (see Paper II):
J033241.88-274853.9 appears to be very asymmetric with a more
elongated side toward North. This tail shows several blue regions with
B-z colors typical of star-bursting galaxies, while the overall color
is representative of late-type galaxies (B-z=1.3, see Paper II). The
reddish edges of J033241.88-274853.9 suggests a moderate presence of
dust in this galaxy.

The stellar light distribution of J033226.23-274222.8 was analyzed
using a two-component Sersic decomposition on the z-band image with
GALFIT \citep{peng02}. We required these two components to be both
centered on the dynamical/continuum center of the galaxy (see below)
and masked the southern part J033241.88-274853.9 to fit only the
northern region in order to be able to reproduce its asymmetry. We
found Sersic indexes n=0.34 and n=0.54 for the outer and inner
components respectively, with an inner-to-outer light ratio
$\sim$0.12. This suggests that J033241.88-274853.9 is composed of an
inner symmetric stellar disk embedded in another bluer extended
``half-disk''.

\begin{figure*}
\centering
\includegraphics[width=18cm]{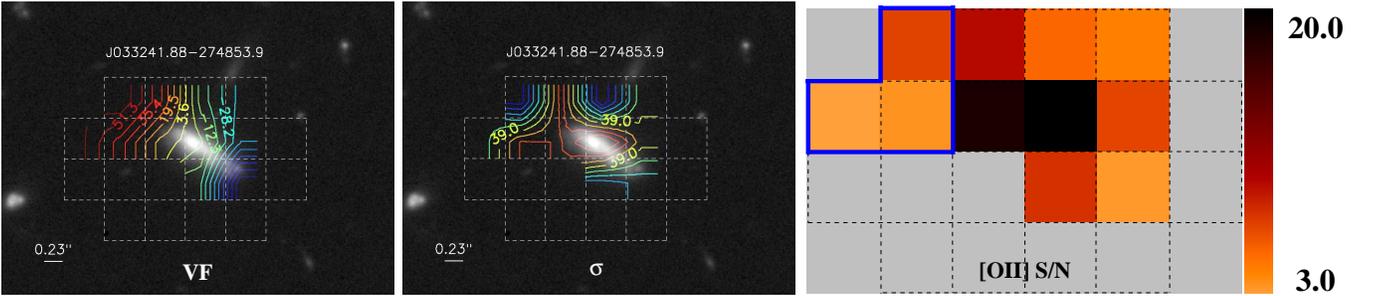}
\caption{B+V+z image of J033226.23-274222.8 superimposed with GIRAFFE
  data. In the left and middle panels, the IFU bundle is overlaid in
  white dash lines (0.52 arcsec/pix), and the residual uncertainty on
  the alignment between the images and the IFU is shown as an
  horizontal white bar in the lower-left corner.\emph{Left panel:}
  isovelocities have been superimposed, with contours approximately
  ranging from -90 to 60 km/s by steps of 8 km/s. A S/N=3 threshold is
  used to limit measurement uncertainties, which explains why
  kinematical data do not extent over the whole IFU FoV. \emph{Middle
  panel:} GIRAFFE velocity dispersion map contours, ranging from 27 to
  47 km/s by steps of 2 km/s. \emph{Right panel:} [OII] emission
  Signal-to-Noise map in GIRAFFE IFU pixels. Pixels where no stellar
  light has been detected in ACS images are ensquared in blue.}
\label{all}
\end{figure*}

In Fig. \ref{all}, we have superimposed the GIRAFFE velocity Field
[VF] and the velocity dispersion-map [$\sigma$-map] to the ACS imaging
(see \citealt{puech07b} for detail). An additional correction was
performed by eye, by aligning the B-band HST/ACS image and the [OII]
emission line distribution (not shown here), since in principle they
both trace the spatial distribution of O/B blue giant stars. Aligning
the peaks of both distributions, we ended up with an additional shift
of 0.2 arcsec in RA and 0.03 arcsec in DEC, i.e., smaller than the
residual uncertainty of the astrometric alignment (0.23 arcsec). No
large-scale perturbation can be seen in the kinematics of
J033241.88-274853.9, which was classified as a rotating disk in Paper
I. The central peak in the $\sigma$-map is well reproduced by a
rotating disk model and indicates the position of the dynamical center
of rotation, which matches relatively well the continuum peak (see
Fig. \ref{all} and Paper I). One can however note that this central
peak of dispersion has two extensions in the southern part, which are
not reproduced by the rotating disk model of Paper I. Moreover, the
gaseous phase in J033241.88-274853.9 has a relatively low
$V/\sigma$=3.79$\pm$0.94 value, which is among the lowest values found
in the sample of distant rotating disks \citep{puech07a}. Such a value
is lower than the local median and, together with the two extensions
of the dispersion peak, it suggests than the ionized gas in the
southern part of J033241.88-274853.9 is relatively turbulent. We will
discuss further the excess of velocity dispersion relative to a
rotating disk model in Sect. 5.

Of importance, GIRAFFE detected [OII] emission on the southern side,
where no stellar emission can be detected, even in the very deep ACS
images of the HUDF (10-$\sigma$ detection limit of $V_{AB}=$29.3
mag/arcsec$^2$ in the 0.52x0.52 arcsec$^2$ GIRAFFE IFU pixel aperture,
see the [OII] S/N map on Fig. 2). Such an asymmetry between gas and
star distributions is unusual. We used a VIMOS spectrum of
J033241.88-274853.9, retrieved from the VVDS (\citealt{lefevre05}, see
Fig. \ref{spec}), and the GIRAFFE [OII] emission map to estimate the
[OII] flux in the leftest GIRAFFE pixel. We degraded the V-band ACS
image down to the spatial resolution and sampling of GIRAFFE
observations to estimate the rest-frame pseudo-continuum around the
[OII] emission line. We were then able to roughly estimate the
rest-frame [OII] equivalent width in this pixel, which was found to be
EW$_0$([OII])=140$\AA$. We will discuss this point in Sect. 5.

\section{Spectral Energy Distribution}
\begin{figure}
\centering
\includegraphics[width=9cm]{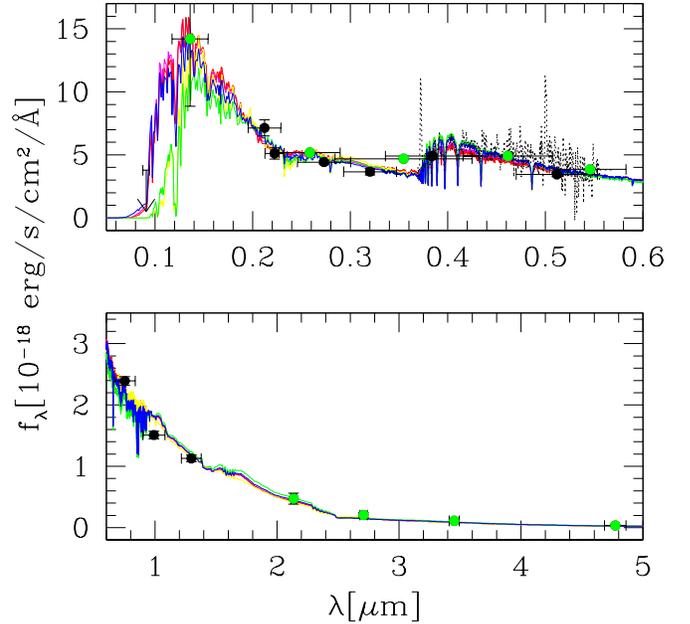}
\caption{Rest-frame SED of J033241.88-274853.9 (3-arcsec fixed
aperture photometry). Green points represent space-based photometry,
while black-points represent ground-based photometry. The GALEX/FUV
5-$\sigma$ detection limit is represented as a vertical arrow. A VIMOS
spectrum is overlaid using a black dashed line (see also Fig.
\ref{spec}). Several stellar population models are shown, all
acceptable within a 90\% confidence interval (see text): a CSP with
$Z=Z_\odot$, $\tau$=0.1 Gyr, $t$=0.32 Gyr, and $A_V=R_V E(B-V)$=0.28
(best fit, blue curve), a SSP with $Z=Z_\odot$, $t$=0.1 Gyr, and
$A_V$=0.07 (green curve), a CSP with $Z=Z_\odot$, $\tau$=0.5 Gyr,
$t$=0.64 Gyr, and $A_V$=0.2 (magenta curve), a SSP with
$Z=2.5Z_\odot$, $t$=0.09 Gyr, and $A_V$=0.7 (yellow curve), and a CSP
with $Z=Z_\odot$, $\tau$=0.35 Gyr, $t$=0.18 Gyr, and $A_V$=0.8 (red
curve).}
\label{sed}
\end{figure}

We gathered photometric measurements of J033241.88-274853.9 in
constant apertures of 3 arcsec in diameter, as shown in Fig \ref{sed}.
We distinguished between space-based measurements, in green, and
ground-based measurements, in black. Space-based measurements
correspond to, from the UV to the IR, GALEX/NUV \citep{morrissey07},
HST/ACS (F435W [B], F606W [V], F775W [i], and F850LP [z]
\citealt{beckwith06}), and Spitzer/IRAC (3.6, 4.5, 5.8, and 8.0
$\mu$m, Dickinson et al. in prep.), while ground-based measurements
correspond to EIS (U', U, B, V, R, and I bands in the optical,
\citealt{arnouts01}), and GOODS-ISAAC (J, H, and K$_S$ bands, Vandame
et al. in prep) in the NIR. Because of the large uncertainty
associated with the zero point calibration of the GOODS H-band, we
used the revised calibration proposed by \cite{wuyts08}. Finally,
J033241.88-274853.9 was neither detected by GALEX in the FUV (the
5-$\sigma$ detection limit is represented as a vertical arrow),
Spitzer/MIPS in the Far-Infrared (Chary et al. in prep.), or CHANDRA
in X-rays \citep{giacconi02,rosati02}.

Full horizontal error-bars represent the FWHM of the photometric
filters used in the corresponding surveys, while vertical error-bars
represent the convolution of two terms, i.e., the magnitude error as
given by Sextractor \citep{bertin96}, and the systematic error
associated with the uncertainty on the zero point calibration. Because
of the very large size of both the GALEX PSF and IRAC PSF in
comparison to the photometric aperture, direct photometric
measurements were corrected by 2.09 mag in the NUV, following
\citep{morrissey07}, and 0.35, 0.38, 0.51, 0.54 mag in the four
respective IRAC IR bands, as directly estimated from the theoretical
IRAC PSF\footnote{http://ssc.spitzer.caltech.edu/irac/psf.html}. We
estimated the uncertainties associated with these PSF corrections by
re-deriving these corrections in apertures of 3 arcsec modulo one PSF
pixel, and added them in quadrature with other sources of uncertainty.
Photometric measurements and uncertainties are listed in Tab.
\ref{tabphot}.

\begin{table*}
\centering
\caption{Photometric measurements for J033241.88-274853.9. \emph{From
    left to right:} 3-arcsec aperture photometry, random errors as
  given by Sextractor, systematic uncertainties associated with the
  calibration of the zero point, and uncertainties associated with the
  PSF-size correction (see text).}
\begin{tabular}{cccccc}\hline\hline
Filter & Magnitude (AB) & Random error & ZP errors & PSF errors & Total uncertainty\\\hline
FUV$_{GALEX}$   & $<$ 25.8 & -- & -- & -- & --\\
NUV$_{GALEX}$   & 23.49 & 0.12 & 0.03 & 0.38 & 0.40\\
U'$_{EIS}$  & 23.27 & 0.05 & 0.083 & 0.0 & 0.10\\
U$_{EIS}$ & 23.53 & 0.05 & 0.083 & 0.0 & 0.10\\
B$_{UDF}$ & 23.19 & 0.004 & 0.005 & 0.0 & 0.006\\
B$_{EIS}$ & 23.24 & 0.02 & 0.029 & 0.0 & 0.04\\
V$_{EIS}$ & 23.10 & 0.04 & 0.066 & 0.0 & 0.08\\
V$_{UDF}$ & 22.61 & 0.0015 & 0.005 & 0.0 & 0.005\\
R$_{EIS}$ & 22.39 & 0.02 & 0.048 & 0.0 & 0.05\\
I$_{EIS}$ & 22.15 & 0.03 & 0.044 & 0.0 & 0.05\\
I$_{UDF}$ & 21.99 & 0.0010 & 0.005 & 0.0 & 0.005\\
z$_{UDF}$ & 21.89 & 0.0016 & 0.005 & 0.0 & 0.005\\
J$_{GOODS}$ & 21.71 & 0.01 & 0.03 & 0.0 & 0.03\\
H$_{GOODS}$ & 21.61 & 0.02 & 0.04 & 0.0 & 0.05\\
K$_{GOODS}$ & 21.34 & 0.02 & 0.05 & 0.0 & 0.05\\
3.6 $\mu$m$_{IRAC}$ & 21.20 & 0.2 & 0.02 & 0.04 & 0.21\\
4.5 $\mu$m$_{IRAC}$ & 21.59 & 0.2 & 0.02 & 0.04 & 0.21\\
5.8 $\mu$m$_{IRAC}$ & 21.73 & 0.2 & 0.02 & 0.03 & 0.20\\
8.0 $\mu$m$_{IRAC}$ & 22.26 & 0.2 & 0.02 & 0.03 & 0.20\\\hline
\end{tabular}
\label{tabphot}
\end{table*}

The resulting SED of J033241.88-274853.9 shows a globally declining
curve from the UV to the IR (see Fig. \ref{sed}). While the optical
part of the SED is roughly compatible with that of an late-type galaxy
(e.g., the observed B-z color, see Sect. 2, which approximately
corresponds to a rest-frame $\sim$2800-V color), its UV part shows an
excess of light that is produced by massive blue stars (O/B), and
certainly indicates a recent burst of star formation in the blue
regions described in Sect. 2 (see Fig. \ref{view}).

We used Charlot \& Bruzual models (Charlot \& Bruzual 2007, in prep;
\citealt{bruzual07}) to fit this SED: we assumed a Salpeter IMF
\citep{salpeter55} and constructed grids of metallicity (0.005, 0.02,
0.2, 0.4, 1.0, 2.5, 5.0 Z$_\odot$), $\tau$-exponentially declining
star formation histories spaced at 0.1 Gyr intervals from $\tau$=0
(pure SSP) to 0.5 Gyr, and at 0.5 Gyr intervals beyond, and age $t$.
Dust was accounted for using a \cite{cardelli89} extinction curve
parametrized by $E(B-V)$ and $R_V$, which was allowed to range between
2 and 6, i.e. typical values found in local group galaxies (e.g.,
\citealt{fitzpatrick07}). For convenience, this extinction curve was
extrapolated in the IR and the FUV. We checked that the results were
not significantly changed assuming another extinction curve such as
the one of \cite{fitzpatrick99} or \cite{calzetti00}.

We searched for the best model using a classical $\chi ^2$
optimization. Several well-known difficulties are associated with this
procedure (e.g., \citealt{papovich01}): (1) the photometric
uncertainties are often under-estimated, (2) additional systematics
arise from mismatches between models and reality (e.g., wrong dust
model, IMF, or star formation history, finite grid of parameters for
the models, etc.), (3) the influence of non-Gaussian errors can bias
the $\chi ^2$ optimization. To check that (1) cannot impact severely
our results, we compared our photometric measurements to those made in
similar bands by \cite{wuyts08} in 2 arcsec apertures, given in their
FIREWORKS catalog. We found that the SED derived from FIREWORKS is, as
expected, shifted toward fainter magnitudes (by roughly $\sim$0.1
mag), and that additional variations are well within the adopted
uncertainties, which means that the adopted error-bars are not too
severely underestimated. Moreover, during the fitting procedure, we
enlarged the total photometric uncertainties by propagating the
uncertainty associated with the center of the filter (i.e., horizontal
error-bars in Fig. \ref{sed}). This allowed us to take into account
the additional uncertainty associated with the projection of the
models onto the photometric filters, which also help mitigate against
(2). Confidence intervals at 68\% and 90\% were determined using
constant $\chi ^2$ bounds, following \cite{avni76}.

The best fit was found for a CSP with $Z=Z_\odot$, $\tau$=0.1 Gyr,
$t$=0.32 Gyr, and $A_V=R_V E(B-V)$=0.13 (see the blue curve on Fig.
\ref{sed}), which corresponds to a reduced $\chi ^2$=1.36. This is
slightly larger that 1 (corresponding to a perfect match) and
certainly reflects the influence of the limitations listed above.
Within a confidence interval of 68\% (i.e., 1-$\sigma$ for a Gaussian
distribution), we found $Z=Z_\odot$, $\tau \in$ [0; 500] Myr, $t \in$
[0.13; 0.64] Myr, and $A_V \in$ [0; 0.28], while for a confidence
interval of 90\%, we found $Z \in$ [1.0; 2.5]$Z_\odot$, $\tau \in$ [0;
3500] Myr, $t \in$ [0.06; 0.91] Myr, and $A_V \in$ [0; 0.74] (see Fig.
\ref{sed}). We note that a solar metallicity is further supported by
the low resolution VIMOS spectrum retrieved from the VVDS archive (see
Fig. \ref{spec}).

We also note the presence of Balmer absorption lines in this best SED
model, which suggests the presence of an underlying older stellar
population (A/F stars). To quantify this further, we tried to fit an
additional stellar population of old stars following the method
described by \cite{papovich01}: we generated SSPs with the same
parameters that the best-fit SED except its age, which was set to be
the age of the Universe at the redshift of the object ($\sim$7.5 Gyr).
In principle, this allows us to constrain the maximal contribution of
an older stellar population in terms of M/L that is still consistent
with the data.

\begin{figure}
\centering
\includegraphics[width=9cm]{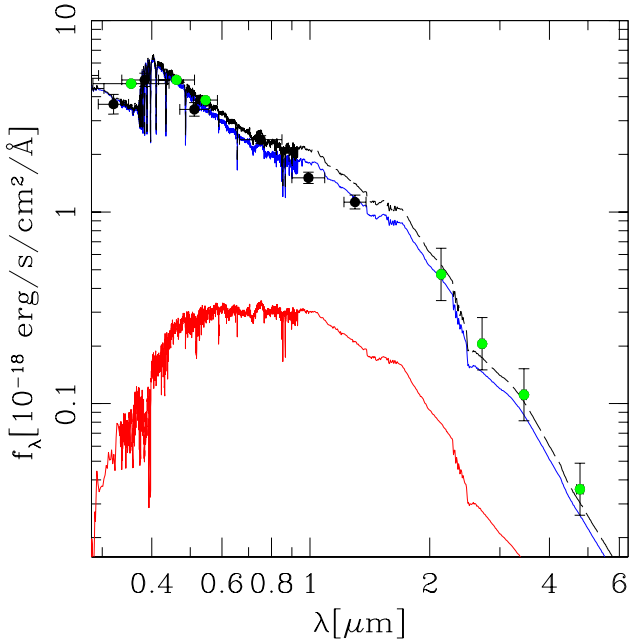}
\caption{Rest-frame SED of J033241.88-274853.9 for wavelengths above
4000\AA. The additional older population fitted together with the
best-fit model (here in blue) is shown in red. The dashed-black line
shows the sum of the two models.}
\label{sedold}
\end{figure}

The result of the fit of this additional population is shown in Fig.
\ref{sedold}. As expected, this additional old population influences
the resulting SED only above 4000\AA, and particularly in the NIR.
One might wonder whether this additional old population is real, given
that the resulting combined SED does not seem to improve the global
fit in this region, where the addition of such an older population, if
real, should in principle lead to an improvement. This is due to the
fact that we have not constrained the best mix between the young and
old populations, but the maximal contribution of old stars to the best
young population model (see above and \citealt{papovich01}), which is
a more conservative approach. To quantify whether or not the addition
of this older stellar population leads to a better SED fitting, we
performed an F-test between the ``old+young'' model and the original
``young'' model. This test reveals that the addition of an older
population does not give a significantly better representation of the
observed SED, as also noticed by \cite{papovich01} in most of their
more distant LBGs.  Actually, constraining an old stellar population
from the SED alone is not an easy task as these stars have a very
modest contribution to the light: it is not surprising that trying to
disentangle old stars from younger ones in an SED does not lead to
conclusive results. Therefore, while there is no direct evidence for
the presence of an older stellar population, its presence cannot be
excluded either.

In conclusion, the SED fitting evidences that J033241.88-274853.9 is a
relatively low-mass object, whose light is little impacted by dust,
and dominated by a very young stellar population aged by a few
hundreds of Myr. These stars appeared to be formed during a relatively
short burst lasting less than a Gyr.

\section{Stellar mass, SFR, and gas fraction}

\subsection{Stellar mass}
Using the best-fit ``young'' model, the stellar mass was found to be
$\log({M_{stellar}/M_\odot})=9.45 \pm ^{0.17} _{0.14}$ (90\%
confidence intervals). This is a factor $\sim$8 smaller than the
stellar mass derived in Paper III, using \cite{bell03} simplified
prescription (see Annex of Paper III for detail), which can be easily
explained by the very young age of J033241.88-274853.9: in this case,
it is well known that using such simplified prescription can lead to
strongly overestimate the stellar mass \citep{bell03}. However, as
discussed above, we cannot exclude the presence of old stars that
could contribute substantially to the global stellar mass, even if
they do not contribute significantly to the light. Considering the two
stellar components model of Fig. \ref{sedold}, we find a total stellar
mass of $\log({M_{stellar}/M_\odot})=9.73$, with a contribution of the
old stellar population of $\log({M_{stellar}/M_\odot})=9.4$, i.e.,
roughly half of the total stellar mass. As the presence of this
population remains hypothetical, we adopt the stellar mass
corresponding to the ``young'' population model, and take this
possible additional contribution into account in the error-bar, which
leads us to $\log({M_{stellar}/M_\odot})=9.45 \pm ^{0.28} _{0.14}$.

\subsection{Star formation rate}
The 100-Myr average SFR was found to be 2.8$^{+3.7} _{-2.8}$M$_\odot$/yr,
in good agreement with the estimate from the rest-frame magnitude at
2800 $\AA$, i.e., SFR$\sim$3.3M$_\odot$/yr. Using the fitted extinction
curve with $A_V$=0.13 mag, we find $A_{2800}$=0.33 mag, which in turn
leads to a dust-corrected SFR of 4.47$M_\odot$/yr. At this SFR, the
mass doubling-time of J033241.88-274853.9 is found to be 0.63 Gyr,
i.e., significantly lower than the age of the Universe at this
redshift. This confirms that J033241.88-274853.9 formed its young
stars quite rapidly, in a rather single burst of star formation as
opposed to a more continuous star formation history. Finally,
following \cite{lefloch05}, we estimated that the detection limit of
MIPS at 24 $\mu$m corresponds to $\sim$10$M_\odot /yr$ at z$\sim$0.65,
which is consistent with the non-detection of J033241.88-274853.9 in
this band (see Sect. 3). We conclude that the SFR of
J033241.88-274853.9 is therefore in the range 4.5-10 M$_\odot$/yr, if
one takes into account both contributions from the UV and the IR.

\subsection{Gas fraction}
It is possible to estimate the gas fraction of J033241.88-274853.9 by
forcing the galaxy to obey the Schmidt-Kennicutt relation between the
star formation surface density and the gas surface density
\citep{kennicutt98}. Indeed, this relation seems to hold for all
morpho-kinematic types (e.g., \citealt{dimatteo07}). We adapted the
method used in \cite{erb06}, to relate the gas mass surface density to
the SFR surface density derived from the UV. Using the dust-corrected
SFR will first provide us with a lower limit on the gas fraction,
since we ignore the contribution from the re-processed light in the
IR. We used the ACS half-light radius (assuming a total radius of two
times the measured half light radius of 3.5 kpc, see Paper II) to
derive the SFR per unit area, which is converted into a gas surface
density using the Schmidt-Kennicutt relation \citep{kennicutt98}. We
then converted this gas mass surface density into a gas mass by
estimating the maximal extent of [OII] emission from GIRAFFE data (see
Fig. \ref{all}), quadratically deconvolved from the effect of a 0.8
arcsec seeing, which led us to $f_{gas}$=73\%. Similarly, we estimated
an upper limit to the gas fraction using the upper limit on the SFR of
10 $M_\odot$/yr, which gives $f_{gas}$=82\%. We conclude that
J033241.88-274853.9 is a very gas-rich galaxy with $f_{gas}$ in the
range 73-82\%. We will discuss this point in the next section.

\section{The J033241.88-274853.9 disk formation process}

The gaseous phase of J033241.88-274853.9 is found to be rotating and
turbulent (see Sect. 2). Because of the relatively coarse spatial
resolution of the GIRAFFE IFU, it is not possible to derive accurately
enough the exact geometry of the gaseous phase from the [OII] map
alone. In this section, we examine two possible geometries for this
gaseous phase, which corresponds to two distinct scenarios for the
formation of the stellar disk.

\subsection{Is J033241.88-274853.9 a collapsing gas-rich rotating disk?}
The kinematical properties of the ionized gas suggests that the
gaseous phase might be distributed into a rotating thick disk. In this
scenario, this pre-existing gas-rich disk would have undergone a starburst
event $\sim$320 Myr ago, partly collapsing into a young stellar disk.

If this case, then this galaxy should be dynamically relaxed and lie
on the Tully-Fisher relation. However, the stellar mass derived from
the SED fitting appears to be $\sim$5 times smaller than the one
predicted by the stellar-mass Tully-Fisher Relation (smTFR) at
z$\sim$0.6 \citep{puech08}. On the one hand, this offset relative to
the distant TFR could suggest that J033241.88-274853.9 is quite
peculiar, as most of the outliers to the distant TFR usually lie
\emph{above} the relation \citep{puech08}. On the other hand, such an
offset from the smTFR has been observed in local dwarf galaxies:
because they have very large gas fractions, they lie below the
smTFR. If one accounts for all baryons, then all galaxies follow the
same ``baryonic'' TFR (bTFR, \citealt{mcgaugh05}). To test if the
large gas fraction of J033241.88-274853.9 (see Sect. 4) can be
responsible for this shift relative to the TFR, we can derived the gas
fraction needed for this galaxy to lie on the bTFR and see whether or
not it is consistent with the range of $f_{gas}$ derived in
Sect. 4. Using the local smTFR as a proxy for the bTFR
\citep{mcgaugh05}, one can estimate the J033241.88-274853.9 gas
fraction to be $\sim$91\%, with a lower bound of $\sim$83\% if ones
takes into account the effect of a possible old stellar
population. This lower limit is roughly compatible with the range of
value derived in Sect. 4. This means that the spectro-photometric
properties of J033241.88-274853.9 are not inconsistent with a
dynamical state of equilibrium, as expected from a relaxed rotating
disk.

One has to explain the decoupling between the ionized gas and star
distributions (see Sect. 2). Many scenarios have been proposed to
explain gas and/or star lopsidedness in local galaxies, but to our
knowledge, none of them appears able to explain a decorrelation
between star and gas distribution such as the one observed in
J033241.88-274853.9 (e.g., \citealt{mapelli08}). In order to check
whether seeing effects and the coarse spatial sampling of GIRAFFE
could explain this apparent decoupling, we used the degraded V-band
ACS image constructed in Sect. 2 to estimate the amount of [OII]
emission that could be spread out by seeing effects in the region
encompassed by the leftmost GIRAFFE pixel. Indeed, assuming that
J033241.88-274853.9 is a relaxed rotating disk, then one expects that
at zero order, the rest-frame B-band (corresponding to the observed
V-band) stellar and ionized gas distributions should be correlated. If
one further assumes that the [OII] equivalent width is spatially
constant, then comparing the GIRAFFE [OII] distribution to the
smoothed ACS V-band image underpredicts the EW$_0$([OII]) in the
leftmost pixel by a factor of 10: we find that the flux ratio between
the region encompassed by the brightest GIRAFFE pixel (i.e., the
GIRAFFE pixel corresponding to the dynamical center of rotation) and
the region encompassed by leftmost GIRAFFE pixel is ten times larger
than the same ratio directly measured in the GIRAFFE [OII] emission
map. This would mean that photo-ionization can account for only
$\sim$10\% of the total EW$_0$([OII]) in this pixel. However, [OII]
and rest-B fluxes trace star formation on different timescales, since
[OII] traces only the youngest stars. Therefore, it could be possible
to have a true [OII] equivalent width that varies spatially. Given the
relatively coarse spatial resolution of GIRAFFE observations, it is
likely that this does not qualitatively change the fact that, if
J033241.88-274853.9 is a relaxed rotating disk, the dominant
ionization source of the gas in its southern part remains to be
explained.

The rotating disk model faces with another difficulty, which is the
extension of the central peak of dispersion in the southern part (see
Sect. 2). Indeed, the morphological analysis reveals no central bar,
and there is no obvious correlation between the velocity dispersion in
these regions and the color map. Therefore, this excess of velocity
dispersion in this region remains also unexplained.

\subsection{Is J033241.88-274853.9 a major merger remnant?}
Another possibility is that the J033241.88-274853.9 gaseous phase is
made of heated spiraling gas re-accreted after the final coalescence
of a merger. Indeed, the merger scenario fits well with the burst
duration and the SFR in this range of stellar mass (i.e., a few
M$_\odot$/yr during a few hundreds of Myr, see, e.g., \citealt{cox08}
and their G0-G0 merger).

In such a scenario, the gas surrounding J033241.88-274853.9 would not
be dynamically relaxed yet, and would rotate too fast for the
underlying stellar mass. Such an effect has been observed in numerical
simulations of mergers (see, e.g., \citealt{puech07a}). Therefore,
this scenario also provides a reasonable explanation for the shift
relative to the smTFR (see also discussion of ``kinematic cooling'' in
\citealt{atkinson08}).

After the final coalescence of the two progenitors, the gas expelled
away in tidal tails can relax by dynamical friction. At least part of
this gas can be subsequently re-accreted and spirals around the merger
remnant, which can mimic the kinematics of a rotating thick disk at
large spatial scales, as shown by \cite{barnes02}. Moreover, in
principle, the large gas fraction of J033241.88-274853.9 is high
enough to allow a disk to be rebuilt following the merger, as shown by
\cite{robertson06}. Such numerical simulation including star formation
have demonstrated how a merger between two equal-mass, gas-rich disks
could allow a stellar disk to reform with an angular momentum
inherited from the orbital momentum of the progenitors
\citep{robertson06}. Indeed, the specific angular momentum of the gas
in J033241.88-274853.9 is found to be j=1120$\pm$255 kpc.km/s, very
close to local gaseous disks (see \citealt{puech07a}).

As we have shown in the previous section, the lack of stars on the
southern side of the galaxy implies that gas cannot be ionized by hot
stars in this region. The low disk V/$\sigma$ and the excess of
velocity dispersion relative to a rotating disk model in this region
(see Sect. 2) both suggest that the disk might be heated by shocks.
These shocks could be produced by collisions between gaseous tidal
tails formed during the merger. Such shocks could ionize the gas in
this region, while also preventing star formation, which provides a
very simple explanation of the decoupling between the gas and stars
distributions. In absence of significant stellar light, gas continuum
emission alone can indeed result in very high equivalent width, i.e.,
in the range 80-370\AA depending on electron density
\citep{osterbrock}: this can largely account for the observed
EW$_0$([OII]) in the southern region of J033241.88-274853.9.

\begin{figure}
\centering
\includegraphics[width=9.cm]{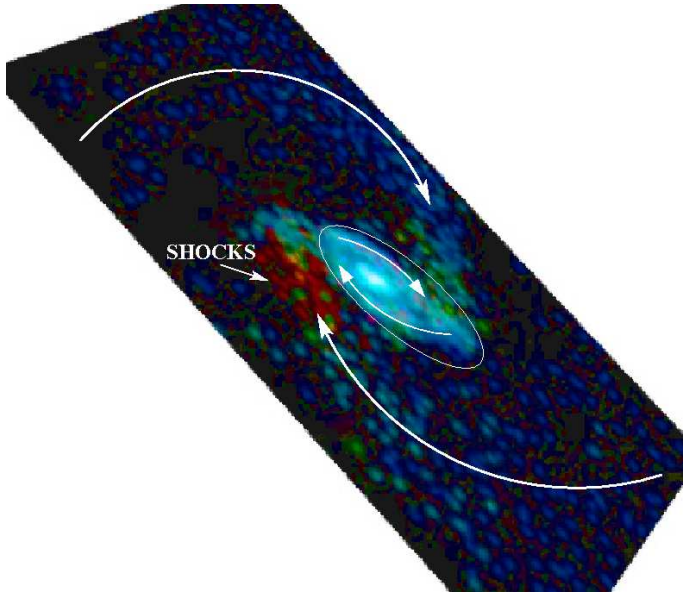}
\caption{Snapshot extracted from a simulation of a major merger by
\cite{barnes02}. Only gas is shown, and colors indicate energy
dissipated by shocks. The view has been rotated and inclined in order
to qualitatively match the morpho-kinematics of J033241.88-274853.9.}
\label{simu}
\end{figure}

In order to test further the merger scenario, we looked in
\cite{barnes02} numerical simulations for similarities with the
properties of J033241.88-274853.9. In Fig. \ref{simu}, we show a
snapshot of the gas distribution extracted from a simulation a major
merger (i.e., mass ratio 1:1) of inclined disks on a parabolic orbit
with a large pericentric separation ($r_{peri}$=0.4, see
\citealt{barnes02} for detail). This snapshot was extracted
approximately 250 Myr after the final coalescence of the two
progenitors, which is consistent with the age of the young stellar
population derived in Sect. 3. Because the stellar light of
J033241.88-274853.9 is dominated by young and intermediate aged stars
(see Sect. 3), the gas concentration can be considered as tracing the
distribution of new stars at very first order, which provides us with
a useful comparison to J033241.88-274853.9. Indeed, in the central
regions, the remnant gas is distributed into an asymmetric rotating
disk (see the white ellipse). The gas expelled during the merger is
re-accreted after the final coalescence through two spiraling tails
(see outer-most white arrows). Shocks are clearly seen at the basis of
the upper tail, which occur between the gas falling down from this
tail onto the central region and the gas rotating around an asymmetric
new disk. The initial gas fraction in the simulation is set to be
12.5\% that of the stellar mass in the disk. It is interesting to note
that in numerical simulations, the rotation support of the re-formed
stellar disk after a major merger is an increasing function of the gas
fraction \citep{robertson06}: with the gas fraction as the one
inferred in J033241.88-274853.9 (see Sect. 4.3), one expects a
strongly rotationally supported remnant stellar disk. The similarities
between J033241.88-274853.9 and this simulation demonstrate the
viability of the merger scenario.

\section{Discussion \& Conclusion}

\begin{table*}
\centering
\begin{tabular}{|c|c|c|}\hline
Property     & Starbursting Rotating Disk & Major Merger Remnant\\\hline
Extension of the peak& & Heating due to shocks\\
of dispersion in the& Cause unknown & between the gas in tidal tails\\
southern part& & and the central regions\\\hline

Decoupling between the & Photo-ionization can account & Shocks can ionize the gas and \\
distribution of & for only 10\%  of $EW_0([OII])$ & account for 100\% of $EW_0([OII])$,\\
the ionized gas \& stars &  &  while preventing star formation\\\hline

\end{tabular}
\caption{Summary of the main characteristics of J033241.88-274853.9. A
possible explanation for each property is given for the two scenarios
discussed in the text.}
\label{tabprop}
\end{table*}

If one assumes that J033241.88-274853.9 is a relaxed rotating disk,
then its gas fraction must be unusually high. Indeed, locally, such
large gas fractions are found only in the most extreme gas-rich dwarfs
galaxies \citep{schombert01}: galaxies on the blue sequence with a
similar stellar mass have gas fraction significantly smaller
\citep{kannappan04}. At z$\sim$0.6, such gas fractions appears to be
also exceptional when compared to the average gas fraction (30\%)
estimated at this redshift by \cite{liang06}. One has to look at
significantly higher redshifts in order to find such large gas
fractions in $\log({M_{stellar}/M_\odot})=9.45$ galaxies (see, e.g.,
\citealt{erb06} for z$\sim$2 galaxies). However, it does not exclude
that we might be observing the collapse of an exceptionally gas-rich
rotating disk into stars.

In this case, one could expect the disk to fragment into clumps, as
described by numerical simulations (e.g., \citealt{immeli04},
\citealt{bournaud07}). In this simulations, gaseous unstable disks can
fragment into UV bright star-forming clumps with typical sizes
$\sim$1kpc. One can note in Fig. \ref{view} the presence of several
blue regions in the disk, which might suggest that the gaseous phase
of J033241.88-274853.9 is indeed fragmented into clumps, as the size
of these regions is found to match that expected in such clumps (see
Fig. \ref{view}). Moreover, this fragmentation process is expected to
occur after 100-300 Myr and to last $\sim$0.5-1.0 Gyr, which might
also fit well with the age inferred above for J033241.88-274853.9.
Such models have also been shown to reproduce the lopsidedness of
distant clump systems, provided that a disk-halo offset is introduced
at the beginning of the simulation \citep{bournaud08}. However, even
with such initial asymmetries, the resulting gas and stars
distributions appear to be relatively well correlated, which is
clearly at odd with the observed decoupling.

\begin{figure}
\centering
\includegraphics[width=9cm]{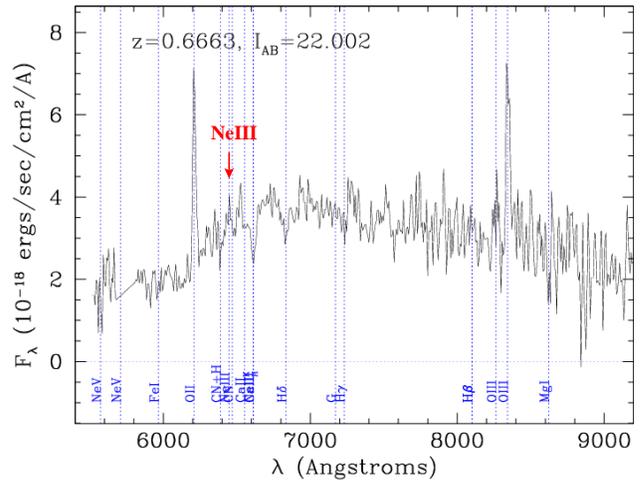}
\caption{VIMOS spectrum of J033241.88-274853.9. We have directly
reproduced the gif figure of this object (EIS ID n$^\circ$24025) as
given by the VVDS database (see text). The expected position of the
NeIII emission line is indicated in red.}
\label{spec}
\end{figure}

Another important argument is the lack of photo-ionization flux in the
southern part of J033241.88-274853.9 to account for the whole
EW$_0$([OII]) measured in this region, while in the major merger
hypothesis, ionization by shocks can easily account for the whole
EW$_0$([OII]). To look for further evidence of the presence of shocks,
we retrieved from the VVDS archive\footnote{http://cencosw.oamp.fr/}
an integrated VIMOS spectrum of J033241.88-274853.9, which is
reproduced in Fig. \ref{spec} \citep{lefevre05}. This spectrum shows
some faint but very likely emission at $\lambda _{rest}$=3869$\AA$,
which could correspond to the NeIII emission line. This strengthens
the presence of ionizing shocks in this galaxy \citep{osterbrock}.

In the previous and present sections, we have identified and discussed
two very unusual properties of J033241.88-274853.9, which are the
decoupling between the distributions of stars and gas, and the excess
of velocity dispersion in the southern part. We have summarized them
in Tab. \ref{tabprop}, along with their respective possible
explanation depending on the two disk formation scenarios considered.
Both properties would require further assumptions to be explained in
the frame of a collapsing gas-rich rotating disk scenario, while they
find a natural and unique explanation through shock-induced ionization
in the frame of the merging of two gas-rich galaxies.

\begin{acknowledgements}
We are grateful to Stephane Charlot who kindly provided us with recent
stellar population models in advance of publication. We thank
J. Barnes for making publicly available his simulations of galaxy
mergers on his web pages
(http://www.ifa.hawaii.edu/\~barnes/research/gassy\_mergers/index.html).
We acknowledge useful discussions with I. Fuentes-Carrera and S.
Peirani regarding the subject of this paper.
\end{acknowledgements}

\end{document}